\documentclass[journal=jacsat,manuscript=article]{achemso}

\usepackage{chemformula} 
\usepackage[T1]{fontenc} 

\author{Zechao Yang}
\email{yangzechao@zedat.fu-berlin.de}
\affiliation[Freie Universität Berlin]
{Fachbereich Physik, Freie Universität Berlin, Arnimallee 14, 14195 Berlin, Germany}
\author{Christian Lotze}
\affiliation[Freie Universität Berlin]
{Fachbereich Physik, Freie Universität Berlin, Arnimallee 14, 14195 Berlin, Germany}
\author{Katharina J. Franke}
\affiliation[Freie Universität Berlin]
{Fachbereich Physik, Freie Universität Berlin, Arnimallee 14, 14195 Berlin, Germany}
\author{Jose I. Pascual}
\email{ji.pascual@nanogune.eu}
\affiliation[CIC nanoGUNE]
{CIC nanoGUNE and Ikerbaske, Basque Foundation for Science, Tolosa Hiribidea 76, 20018 Donostia San Sebastian, Spain}

\title[An \textsf{achemso} demo]
  {Metal-Organic Superlattices Induced by Long-Range Repulsive Interactions on a Metal Surface}

\abbreviations{IR,NMR,UV}
\keywords{American Chemical Society, \LaTeX}

\begin{document}

\begin{tocentry}

  \includegraphics[height=4.5cm]{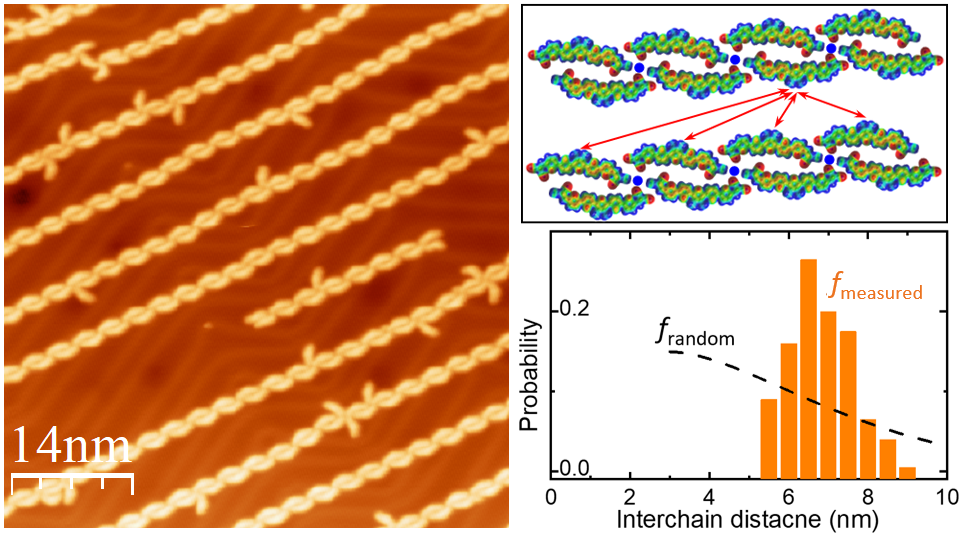}

\end{tocentry}

\begin{abstract}
Chains of Na atoms and dicyanovinyl-quinquethiophene (DCV5T-Me$_2$) molecules with ionic bonds form a superlattice on Au(111). Through a detailed analysis of the interchain distances obtained from scanning tunneling microscopy images at various molecular coverages, we found that the chain arrangement substantially deviates from a random distribution of noninteracting chains. Instead, the distribution of chain spacings provides evidence for the existence of an interchain long-range repulsive potential. Furthermore, the experimental results can be modeled by a repulsive potential with a 1/d distance dependence, characteristic of Coulomb interactions. Density functional theory calculations of free-standing molecular chains reveal a charge depletion at the periphery of the chains, which is attributed to the intrinsic polar property of the molecule and is responsible for the long-range repulsion. 
\end{abstract}

\newpage

The formation of self-ordered organic nanostructures on metal surfaces is driven by intermolecular and substrate-mediated interactions. \cite{doi:10.1146/annurev.physchem.56.092503.141259, Kuehnle2009, Auwaerter2015, Barth2009, Lin2009, Otero2011, Liang2009} By tuning their complex balance, the formation and shape of molecular assemblies can be controlled. \cite{doi:10.1146/annurev.physchem.56.092503.141259, Kuehnle2009, Auwaerter2015, Barth2009, Lin2009, Otero2011, Liang2009} While attractive interactions, such as van der Waals force, hydrogen bonding, metal-organic coordination and ionic bonding, drive the formation of clusters, islands and films, \cite{doi:10.1146/annurev.physchem.56.092503.141259, Kuehnle2009, Auwaerter2015, Barth2009, Lin2009, Otero2011, Liang2009}  intermolecular and surface-mediated repulsive interactions play a significant role in the formation of long-range ordered superlattices. For example, surface-state electrons can support long-range interactions of neutral atoms and molecules at metal substrates. \cite{PhysRevB.65.115420} Alternatively, the repulsion between charges and intrinsic dipoles mediates long-range order of polar and neutral molecules on bulk insulating surfaces. \cite{supramelecularchainsinsulatingsurface} It is reported that the electrostatic repulsive potentials can be preserved on metal substrates. \cite{PhysRevLett.99.176103} As a result, adsorbed charged or dipolar molecules form superlattices of monomers with coverage-dependent molecular pair distributions. \cite{PhysRevLett.98.206102, PhysRevLett.99.176103, orderedmoleculeschargedandsurfaceeffect}

Compared to superlattices of individual adsorbates, the fabrication of one-dimensional arrays of molecular chains with controllable spacing, so-called nanogratings, is more challenging, because it requires an equilibrium between short-range attractive and long-range repulsive interactions. While surface reconstructions can facilitate the formation of nanogratings, their spacing is not tunable but locked to the periodicity of the reconstruction. \cite{Kunkel2015, supramoleculargratingsurfacedrived2, supramoleculargratingsurfacedrived3} Instead, intrinsic interchain repulsive interactions have been used to build nanogratings with controllable distance by tuning the surface coverage. \cite{supramoleculargratingrepulsionpositivchargechainsouterside} In order to obtain robust nanograting structures, it is essential to balance the intrachain attractive forces and the interchain repulsive interactions. Too strong intrachain interaction inhibits the self-ordering and frequently results in curved chains, which happens for covalent wires. \cite{notorderedcovalentchains} On the other hand, if the intrachain interaction is too weak, the chains may break at high coverages induced by interchain repulsive forces, such as chains formed through hydrogen bonding. \cite{supramoleculargratingrepulsionpositivechargeouternotstabehighcoverage} In contrast to covalent bonding and hydrogen bonding, ionic bonding exhibits an intermediate bond strength and is a good candidate for forming chain structures of nanogratings on surfaces. Recently, ionic-bonded compact films on metal substrates were studied. \cite{Skomski2012, C1CC12519B, 1367-2630-15-8-083048} However, to the best of our knowledge, so far nanogratings using ionic bonds to stabilize its chain structure have not been reported.

Here we report the self-assembly of tunable arrays of ionic-bonded metal-organic chains upon co-deposition of sodium chloride (NaCl) and dicyanovinyl-quinquethiophene (DCV5T-Me$_2$) on a Au(111) substrate. By using a combination of low-temperature scanning tunneling microscopy (STM) and density functional theory (DFT), we find that the metal-organic nanograting is formed by the buildup of repulsive long-range interactions between the chains. The coverage-dependent interchain interaction potential landscape was reconstructed through the analysis of interchain distance distributions, and found to be consistent with electrostatic repulsive interactions having a 1/d distance dependence.

\begin{figure*}
\centering
  \includegraphics[height=8cm]{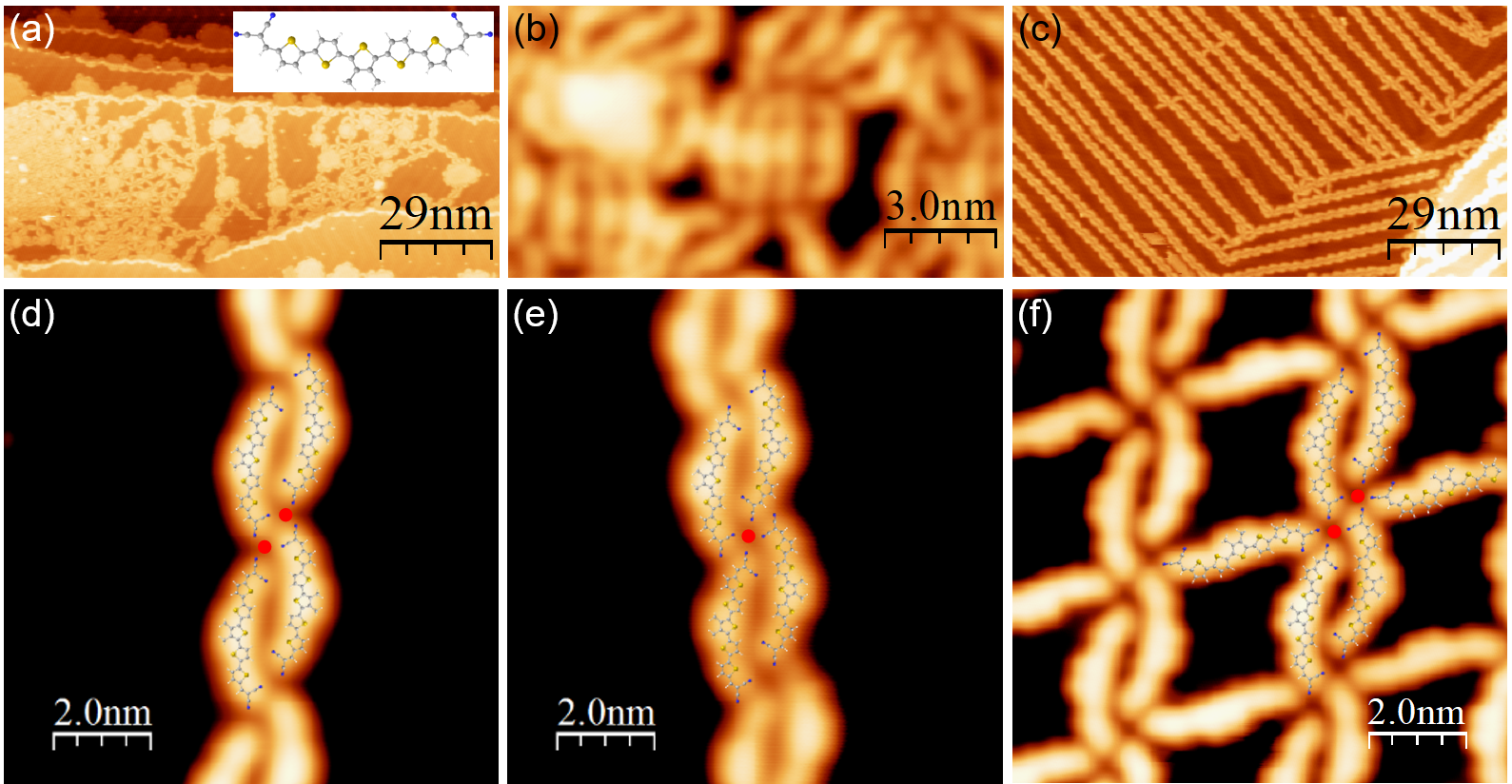}
  \caption{Ionic-bonded chains formed by mixing DCV5T-Me$_2$ with NaCl on Au(111).  (a) STM overview (I = 82 pA, V = 0.65 V) of the co-deposition of DCV5T-Me$_2$ with NaCl at room temperature. (b) High resolution STM image (I = 170 pA, V = 0.65V) of the disordered mixture. (c) STM image (I = 63 pA, V = 0.61 V) after annealing the sample to 400 K. The bonding structure varies depending on the coverage. (d) At low coverage of 0.21 ML (I = 170 pA, V = 0.65 V) the Na-CN bonding frame is three-fold. (e) At intermediate coverage of 0.38 ML (I = 110 pA, V = 0.6 V) the Na-CN bonding frame is four-fold. (f) At high coverage of 0.75 ML (I = 180 pA, V = 0.63 V) porous networks are formed and the four-fold Na-CN bonding frame is maintained. The model of the molecules is superimposed to identify the bonding structure. In (f) one DCV group and one thiophene ring are rotated in the molecules connecting the parallel chains. The red circles represent the Na atoms.}
  \label{fgr:example}
\end{figure*}

DCV5T-Me$_2$ has two dicyanovinyl (DCV) functional groups that are linked symmetrically to a central quinquethiophene (5T) moiety with the central thiophene ring substituted by two methyl (Me$_2$) groups (inset in Figure 1a). The interest in this species lies in the potential for organic solar cells, since it combines electron donor (5T) and acceptor (DCV) moieties within the same organic backbone. \cite{doi:10.1021/jp970085z, CPHC:CPHC200300963, ADMA:ADMA200600658, PhysRevB.77.085311, 5392759, AENM:AENM201100026, ADFM:ADFM201001639, doi:10.1021/cm201392c, ADMA:ADMA201104439} On Au(111), DCV5T-Me$_2$ molecules were found to form either homomolecular islands stabilized by hydrogen bonding \cite{Bogner2015} or metal-organic chains by incorporation of Au adatoms to the bonding motif. \cite{Yang2014} An alternative approach to obtain mixed architectures with homogeneous bonding motif is the use of Na atoms as cation. Na ions are prone to electron donation and create robust ionic phases, mediated by charge-transfer interactions. \cite{Skomski2012} In particular, Na ions, extracted from NaCl layers, were found to effectively bind to CN groups of TCNQ and stabilize two-dimensional charge-transfer layers. \cite{C1CC12519B, 1367-2630-15-8-083048} The same bonding scenario is explored here by co-depositing DCV5T-Me$_2$ and NaCl on Au(111). 

In our experiment, DCV5T-Me$_2$ molecules and NaCl are successively evaporated from a Knudsen evaporator onto an atomically clean Au(111) substrate at room temperature under ultra-high vacuum. All measurements were performed in a low-temperature STM, at a temperature of 5 K. Na atoms and DCV5T-Me$_2$ start to mix when NaCl is deposited onto the pre-formed DCV5T-Me$_2$ assemblies, as shown in Figure 1a. The organic islands are transformed into disordered phases, \cite{Peyrot2017} with cyano groups pinned around small clusters of, presumably, NaCl (Figure 1b). Post-deposition annealing turns out to be essential to transform the disordered phase into the nanogratings covering the whole surface (Figure 1c).

A closer look into the molecular chain structure reveals that the Na-CN ionic bonding at the molecular joints exhibits diverse configurations, ranging from three-fold to four-fold geometry. \cite{Peyrot2017, Nathreefold} We found that the precise domain structure depends on molecular coverage. At low coverage, e.g. 0.21 monolayer (ML) in Figure 1d, the Na-CN bonding is three-fold and two Na atoms are incorporated in every node of the bi-molecular chain (Figure 1d). At the intermediate coverage of 0.38 ML, we find domains with a four-fold Na-CN bonding motif with one Na atom at every node (Figure 1e). We attribute this transformation to the ratio between molecules and Na atoms available for bonding. We assume that at a certain annealing temperature there is an equilibrium between Na atoms confined in NaCl islands and free to bond with ligands. With the increase of molecular coverage, Na atoms available for coordination are not enough for the three-fold bonding geometry, where the ratio between Na and DCV5T-Me$_2$ is 1:1. Hence, the bonding configuration adjusts itself to the available Na atoms by adopting a Na/DCV5T-Me$_2$ ratio of 1:2. This explanation is confirmed by the fact that at higher coverage (0.75 ML) porous Na-DCV5T-Me$_2$ networks are formed with the four-fold Na-CN bonding configuration, as shown in Figure 1f.

\begin{figure*}
 \centering
 \includegraphics[height=8.5cm]{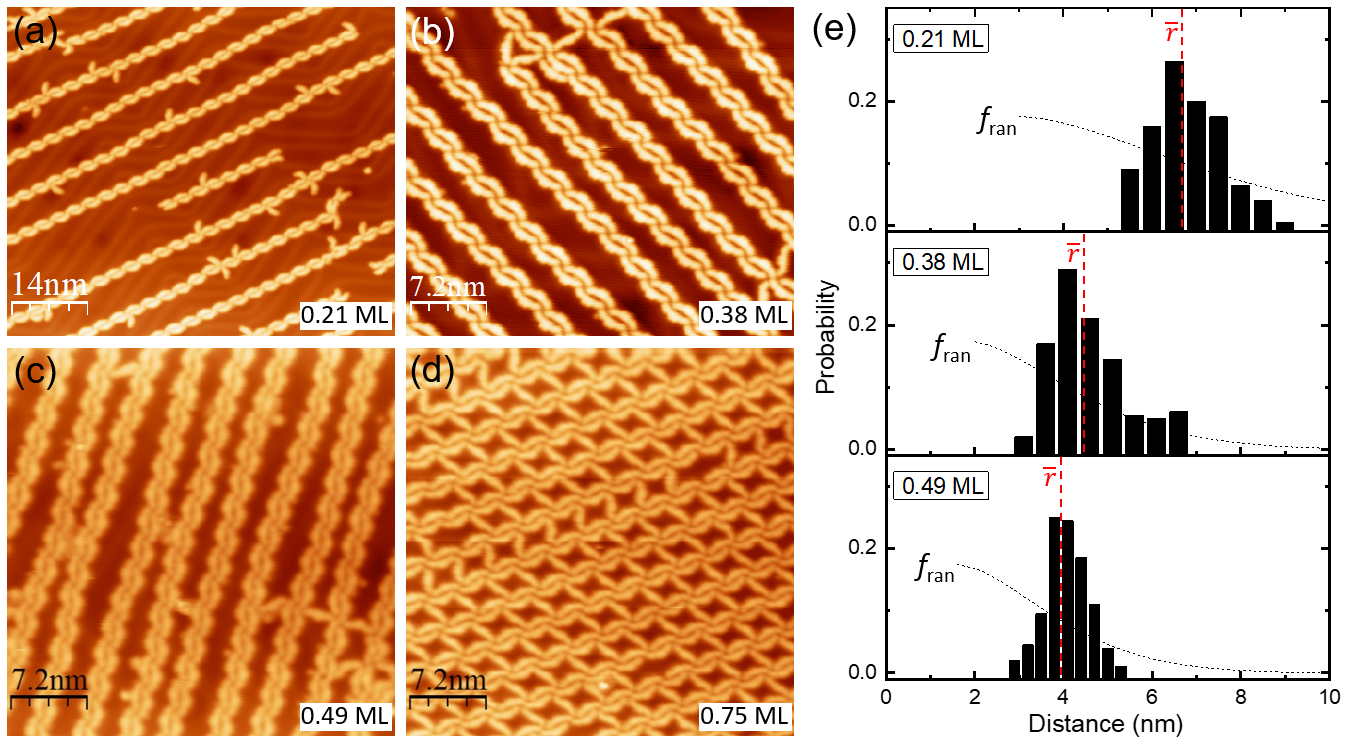}
 \caption{(a)-(d) STM images of DCV5T-Me$_2$/Na chains and networks at various coverages on Au(111). (a) 0.21 ML (I = 100 pA, V = 0.65 V), (b) 0.38 ML (I = 66 pA, V = 0.64 V), (c) 0.49 ML (I = 57 pA, V = 0.63 V) and (d) 0.75 ML (I = 210 pA, V = 0.69 V). The distances between neighboring chains decrease with the increases of coverage, while the chains stay nearly parallel and equidistant. (e) Interchain-distance distribution between adjacent chains at different coverages. The dominant chain distances, $\overline{r}$, are highlighted with red dashed lines (6.8 nm, 4.5 nm and 4.0 nm). The corresponding one-dimensional random distributions for non-interaction chains, $f$$_\mathrm{ran}$, are shown as black dashed lines.}
 \label{fgr:example2col}
\end{figure*}

Independently of their ionic bonding motif, the molecular chains arranged into quasi-periodic one-dimensional superlattices for all explored intermediate coverages. In Figure 2 we compare the results of three different coverages: 0.21 ML, 0.38 ML and 0.49 ML. STM images reveal that the chains stay separated, parallel and equi-distant, with a monotonous decrease of interchain distance with increasing coverage. Porous networks are formed when the coverage reaches 0.75 ML, as shown in Figure 2d. At this coverage, the interchain separation is reduced to the length of one molecule, where the excess molecules connect to the joints of two adjacent chains rather than forming additional chains. The formation of quasi-periodic one-dimensional superlattices, rather than compact assemblies, is a fingerprint of the existence of long-range repulsive interaction.
 
In order to get an insight into the origin of this one-dimensional ionic-bonded superlattice, we performed a statistical analysis of the interchain distance for the three different coverages investigated in Figure 2a-c. \cite{PhysRevLett.99.176103, supramelecularchainsinsulatingsurface} Since the interchain distance is fixed by the molecular length in the porous networks, the coverage of 0.75 ML is excluded in the statistics. The pair-distance distributions are shown in Figure 2e. For each plot 200 chain-pairs are taken into account. The distributions are peaked to interchain distance of 6.8 nm, 4.5 nm and 4.0 nm, from the lowest to the highest coverage, respectively. The one-dimensional random distribution, $f$$_\mathrm{ran}$, corresponding to chains without interactions and plotted for each coverage in Figure 2e as a reference, is expressed as: \cite{PhysRevLett.99.176103, PhysRevB.65.115420} 
\begin{equation}
	f_\mathrm{ran}=\frac{Nar}{L}(1-\frac{2r}{L})^N
\end{equation}
where $N$ is the number of chains per image, $L$ is the distance from the first to the last chain in an image, $a$ is the minimum width of the box that can fit one unit cell of chains inside, and $r$ is the distance between two adjacent chains. In contrast to the broad random distribution profile, the experimental data exhibits a peaked distribution at a certain, most probable pair spacing, which reduces with coverage. Such peaked distributions thus reveal the existence of a repulsive interchain interaction.

There are several mechanisms that can induce the long-range quasi-periodic growth of molecules on metal surfaces. For example, the fcc regions of the Au(111) reconstruction can be preferential adsorption sites for molecules due to the lower concentration of surface atoms, when compared to the buried layers, \cite{PhysRevLett.99.176103, PhysRevB.42.9307} thus forming a one-dimensional super-lattice. However, STM images can rule out this mechanism here because the chains extend across the intact herringbone domain lines rather than parallel to them (Figure 2a). Another possibility is the interaction of chains via a Friedel-like potential, oscillating with half of the Fermi wavelength ($\lambda$$_\mathrm{F}$), which amounts to 1.8 nm for Au(111). \cite{0953-8984-12-1-103, PhysRevB.65.115420, PhysRevLett.88.028301, PhysRevLett.92.016101, PhysRevLett.85.2981, 0953-8984-12-1-103} However, the experimental pair-distance distribution shows much larger interchain separations and non-integer multiples of $\lambda$$_\mathrm{F}$/2. Therefore, the surface-electron-mediated growth can also be excluded here.

The existence of an intrinsic molecular dipole can also give rise to electrostatic intermolecular repulsive interactions. \cite{PhysRevLett.99.176103, PhysRevLett.98.206102} Dipoles within individual DCV5T-Me$_2$ are expected due to the built-in acceptor-donor-acceptor structure. Such electrostatic dipolar nature of DCV5T-Me$_2$ is expected to become even stronger after ionic bonding to Na atoms due to charge transfer and charge reorganization, thus suggesting electrostatic forces as probably responsible for the long-range repulsion between chains.

In order to get a deeper insight into the electrostatic interaction within the molecular chains, we performed density functional theory (DFT) simulations of free-standing Na-DCV5T-Me$_2$ chains. The calculations were performed by using the GAUSSIAN 09 program package with the B3LYP exchange-correlation functional and 6-31G basis set \cite{g09}. As shown before, \cite{Yang2014, Yang2019} the Au(111) substrate plays a negligible role on the bonding geometry and electronic states of the dicyanovinylene-substituted oligothiophenes and their assemblies. Therefore, the calculations were carried out on isolated systems without considering the Au(111) substrate.  

Chain structures are difficult to model in gas phase due to the high degree of freedom induced by the bending and twisting of the chains into three-dimensional configurations. Thus, we focus on studying the ionic bonding by simulating one Na atom placed close to a single DCV5T-Me$_2$ molecule. The relaxed structure has a bond length of 2.1 Å between the cyano group and the Na atom, with the DCV5T-Me$_2$ backbone maintaining a planar configuration, as shown in Figure 3a. The simulations find that 0.817 e$^-$ are transferred from Na to DCV5T-Me$_2$, in agreement with the larger electronegativity of the organic molecule. As reported before, \cite{C1CC12519B, 1367-2630-15-8-083048} this transferred charge is mainly localized around the dicyano groups, while the sulfur atoms and hydrogen atoms remain positively charged (Figure 3b).  	

\begin{figure}[h]
\centering
  \includegraphics[height=9.45cm]{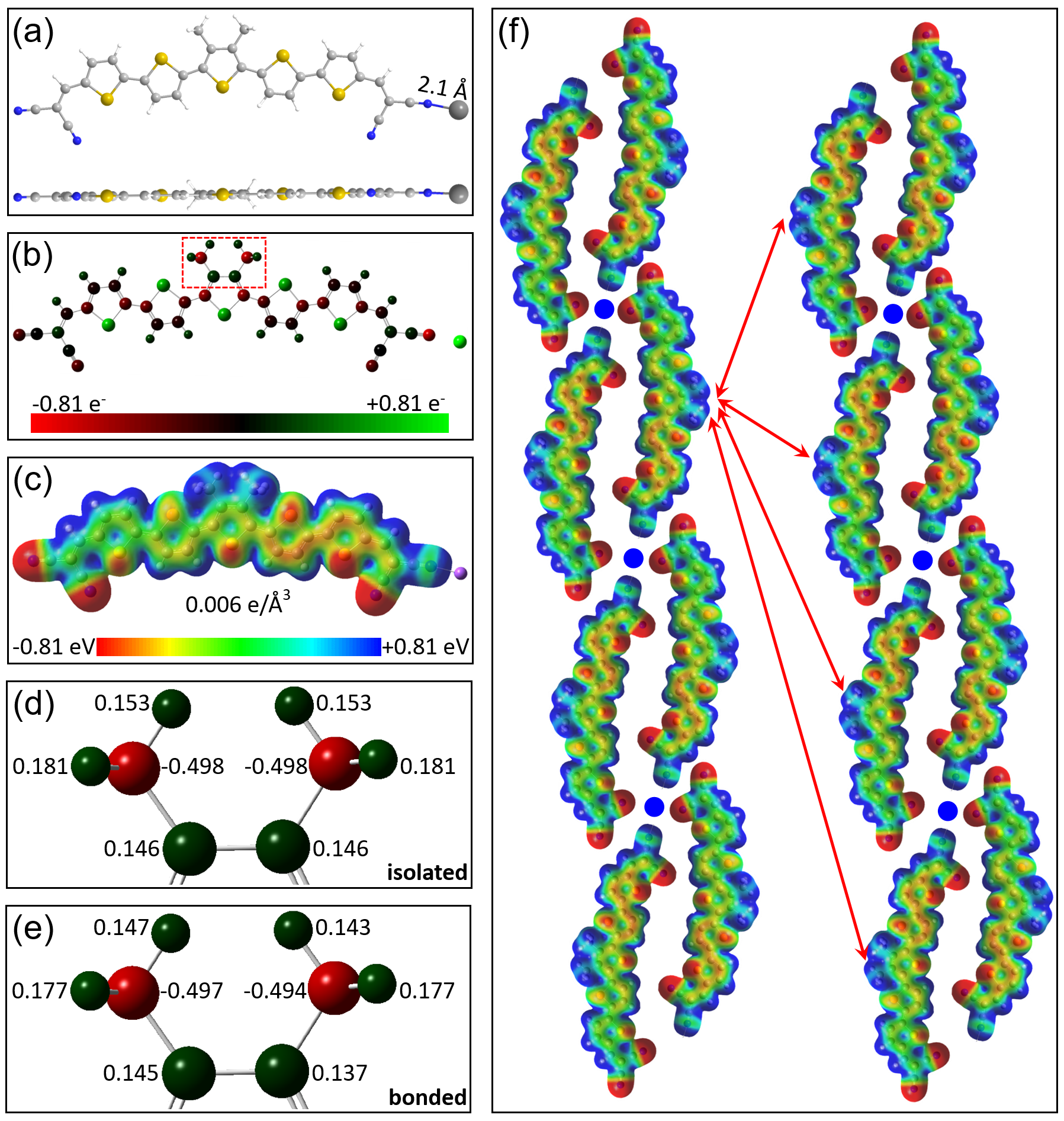}
  \caption{DFT calculations of the Na/DCV5T-Me$_2$ complex in gas phase. (a) Relaxed bonding structure. The gray circle represents the Na atom. (b) Total charge distribution from Mulliken population analysis. (c) Calculated electrostatic potential energy of an iso-surface with total electron density of 0.006 e per cubic angstrom at a molecule with one Na bonded. The red and blue stand for negative and positive potential regions, respectively. (d) Charge distribution at the central methyl groups of an isolated molecule. (e) Charge distribution at the central methyl groups of a molecule bonded with a Na atom. The position of the methyl groups at the molecule is highlighted by the dashed red rectangle in b. The numbers in d and e state the net charge at each atom in the unit of one electron. (f) Model showing the long-range repulsive force between adjacent chains.}
  \label{fgr:example}
\end{figure}
	
Since this intramolecular charge distribution is probably the cause of the repulsive interactions between the chains, we study the electrostatic potential landscape across the molecules. We first generate an iso-surface of constant total electron density from the self-consistent field (SCF) density matrix. Then, we use this iso-surface to map the SCF-generated electrostatic potential energy felt by an electron at a constant distance from the molecular backbone, as shown in Figure 3c. The map reveals that positive potential regions (blue color) caused by electron depletion are localized at the convex part of the molecule, while the electronegative CN end groups appear as negative potential regions (red color). As a result, the assembled chain exposes regions with electron depletion towards the outside, which probably are responsible for the repulsive interchain interactions (as sketched in Figure 3f) that drive their arrangement into one-dimensional superlattices.

A close inspection of the charge redistribution induced by the charge transfer reveals that the ionic bonding has only a very local effect. As shown in Figure 3b and 3c electrons from the Na atom are mainly transferred to the adjacent N atom, \cite{1367-2630-15-8-083048} while the rest of the molecule including the central Me$_2$ moiety remains in a charge state similar to the isolated molecule (Figure 3d and 3e). Thus, our simulations suggest that the ionic charge transfer plays a small role on the long-range repulsion interaction, but, instead, is responsible for the formation of flexible chains that can accommodate their mutual spacing to the minima of the electrostatic potential landscape.

\begin{figure}[h]
\centering
  \includegraphics[height=6.5cm]{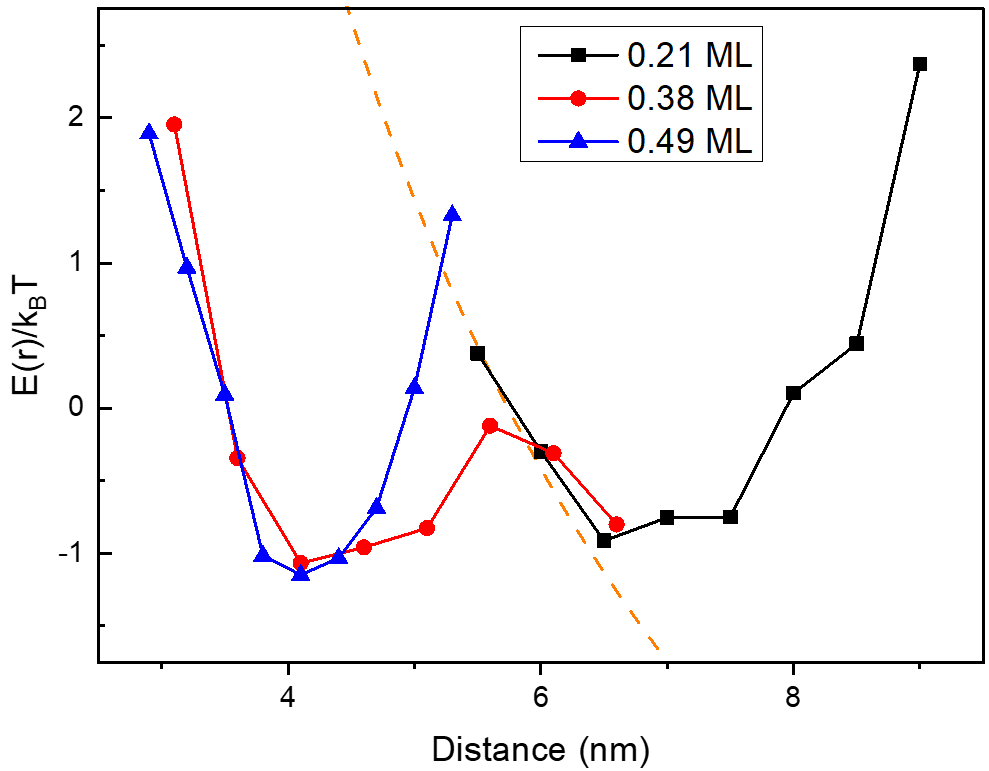}
  \caption{Potential energy between adjacent chains as a function of distance at different coverages. The repulsive forces between chains induce the well shape of the potential and the corresponding dominant average distances (6.8 nm, 4.5 nm and 4.0 nm). The orange dashed line represents the modeled Coulomb repulsion potential energy, acquired by assuming that the interchain interaction is dominated by CH$_3$ charged with one e$^-$. It fits with the potential from statistics in the dilute limit (0.21 ML).}
  \label{fgr:example}
\end{figure}

The interaction potential landscape can be obtained experimentally by further analysis of the statistics of pair distances in Figure 2e. The distribution $f$$_\mathrm{exp}$ and the random distribution $f$$_\mathrm{ran}$ are related to each other by the Boltzmann factor \cite{PhysRevLett.99.176103, PhysRevB.65.115420, PhysRevLett.98.206102}: 
\begin{equation}
	f_\mathrm{exp}=f_\mathrm{ran}\mathrm{exp}[-E(r)/\mathrm{k}_\mathrm{B}T]
\end{equation}
where $E(r)$ is the interaction potential energy as a function of the interchain distance, k$_\mathrm{B}$ is the Boltzmann constant, and $T$ is the temperature of the system. Then the interaction potential energy can be expressed as:
\begin{equation}
	E(r)=-\mathrm{k}_\mathrm{B}T\mathrm{ln}[f_\mathrm{exp}/f_\mathrm{ran}]
\end{equation}
The plot of $-\mathrm{ln}[f_\mathrm{exp}/f_\mathrm{ran}]$, shown in Figure 4, exhibits a well shape for all the three coverages, reflecting the shape of the one-dimensional potential energy landscape producing the superlattice. The potential well becomes deeper and more symmetric with increasing coverage, indicating that the chains are gradually more confined by the increasing repulsive interactions between chains.

In the limit of a dilute system, the potential obtained from the analysis is expected to be a good approximation to the repulsive pair-interaction potential. A recent theoretical study revealed that neutral molecules containing internal dipoles are unable to induce a long-range pattern of image charges. \cite{noimagecharge} Therefore, we can estimate the interaction potential energy of our system by using a simple one-dimensional Coulomb repulsion model. In this model, we assume that the molecules behave as point charges during nucleation and consider the Coulomb potential between a single DCV5T-Me$_2$ molecule and an infinitely long chain \cite{supramelecularchainsinsulatingsurface} at a distance $r$. 

Our DFT calculations show that the six hydrogen atoms at the central methyl groups accumulate a total of one positive net charge. So, if we assume that point charges amounting to +1 e$^-$ are located at the outer part of the molecular chains, the corresponding Coulomb repulsive potential reproduces the potential derived from the pair-distance statistics for the most dilute case (orange curve in Figure 4). This indicates that the methyl groups of the molecules can be treated as positive point charges and are responsible for the repulsive interaction between the chains. The interaction mechanism derived from Figure 4 suggests that the surface electrons on the Au(111) cannot completely screen the charges localized at the molecular building blocks. This can be explained by the large screening length of the Au(111) surface states. \cite{PhysRevLett.99.176103} 

So now we can conclude that the essential factor for the formation of the superlattice is the existence of methyl groups at the central thiophene group. As shown in Figure 3c, the methyl groups favour larger positive charge accumulation that turns out to be the key for the interchain repulsion. To support this, we compare these results with a previous study on dicyanovinyl-sexithiophene (DCV6T) molecule on Au(111). The DCV6T has a similar chemical structure than the DCV5T-Me$_2$, but with no substitution of methyl groups. Our former study found that the organic DCV6T chains predominantly grow at the fcc area of the herringbone reconstruction on Au(111) and can combine into islands, without showing repulsive interactions. \cite{Bogner2016} Furthermore, uniform metal-organic DCV6T chains can be formed by bonding with Co atoms. However, there the growth of the chains is only mediated by the herringbone reconstruction without a hint of long-range repulsive interaction. \cite{yangthesis} Moreover, the ionic metal-organic bonding is another crucial factor for the formation of nanogratings, which stabilizes the chain structure itself by appropriate intrachain attractive interactions.

In summary, here we report the formation of superlattices of metal-organic chains of ionically bonded DCV5T-Me$_2$ molecules and Na on Au(111), whose interchains separation can be tuned by the molecular coverage. A detailed statistical analysis and DFT calculations attribute the long-range ordering to Coulomb repulsion among localized positive charges at the methyl groups of the outer rims of the chains. We speculate that the observed mechanism can be utilized to fabricate designed nanostructures, like nanogratings, on metal surfaces.

\section{EXPERIMENTAL METHODS}

\subsubsection*{Experiment}  
The experiments were carried out in a home-built low-temperature STM working at a temperature of 5 K and under ultra‐high vacuum (UHV) conditions. The Au(111) surface was cleaned \textit{in situ} by repeated cycles of Ne$^+$ sputtering and subsequent annealing to 750 K. The DCV5T-Me$_2$  molecule was evaporated onto the cleaned sample held at room temperature \textit{in situ} in the UHV chamber from an organic molecular evaporator at 510 K with a quartz balance to control the deposition coverage. The NaCl molecule, source of Na atoms to mix with organic molecules, was evaporated onto the sample in the same way at 800 K. We tried to evaporate the DCV5T-Me$_2$ and the NaCl in different order. Nevertheless, post-annealing to 400 K results in the same metal-organic chain structure on the surface. The STM topographic images were acquired in constant current mode. The indicated bias voltages refer to the sample. The topographic data were processed with Nanotec WSxM software.\cite{:/content/aip/journal/rsi/78/1/10.1063/1.2432410}

\subsubsection*{Theory}  
DFT calculations were performed using the GAUSSIAN 09 program package.\cite{g09} Calculations including the molecular geometry optimization and the ionic bonding were carried out in gas phase using the B3LYP exchange‐correlation functional and the 6‐31G basis set. In the calculation, the geometry of an isolated DCV5T-Me$_2$ molecule was optimized, which is followed by placing a Na atom close to the optimized molecule for the ionic bonding calculation.

\begin{acknowledgement}

The authors thank the group of Prof. Dr. Peter Bäuerle and Dr. Elena Mena from Universität Ulm for providing the DCV5T-Me$_2$ molecules. Funding by the SFB 951 “Hybrid Inorganic/Organic Systems for Optoelectronics” (project 182087777; K.J.F.) is gratefully acknowledged.

\end{acknowledgement}


\bibliography{chains}

\end{document}